\documentclass[twocolumn]{article}
\usepackage[english]{babel}

\usepackage[letterpaper,top=2cm,bottom=2cm,left=3cm,right=3cm,marginparwidth=1.75cm]{geometry}

\usepackage{amsmath, amssymb} 
\usepackage{caption}
\usepackage{subcaption}
\usepackage{graphicx}
\usepackage[colorlinks=true, allcolors=blue]{hyperref}

\title{\textit{nekCRF}: A next generation high-order reactive low Mach flow solver for direct numerical simulations}

\author{
Stefan Kerkemeier$^{a}$, 
Christos E. Frouzakis$^{b}$,  
Ananias G. Tomboulides$^c$, \\
Paul Fischer$^d$, 
Mathis Bode$^{e,*}$
\\
{\footnotesize \em $^a$Micodat Computer GmbH, Germany} \\[-5pt]
{\footnotesize \em $^b$Combustion, Acoustics \& Flow Physics Laboratory, ETH Z\"{u}rich, 8092 Z\"{u}rich, Switzerland}  \\[-5pt]
{\footnotesize \em $^c$Department of Mechanical Engineering, Aristotle University of Thessaloniki, Greece}  \\[-5pt]
{\footnotesize \em $^d$University of Illinois Urbana-Champaign, U.S.A.}\\[-5pt]
{\footnotesize \em $^e$J\"{u}lich Supercomputing Centre, Forschungszentrum J\"{u}lich GmbH, 52425 J\"{u}lich, Germany}  \\[-5pt]
{\footnotesize \em $^*$corresponding author, email: 
\href{mailto:m.bode@fz-juelich.de}{m.bode@fz-juelich.de}}  \\[-5pt]
}

\date{}

\begin{document}
\maketitle

\begin{abstract}

Exascale computing enables high-fidelity simulations of chemically reactive flows in practical geometries and conditions, and paves the way for valuable insights that can optimize combustion processes, ultimately reducing emissions and improving fuel combustion efficiency. However, this requires software that can fully leverage the capabilities of current high performance computing systems. The paper introduces \textit{nekCRF}, a high-order reactive low Mach flow solver specifically designed for this purpose. Its capabilities and efficiency are showcased on the pre-exascale system JUWELS Booster, a GPU-based supercomputer at the J\"{u}lich Supercomputing Centre including a validation across diverse cases of varying complexity.

\end{abstract}

\section{Introduction}

The urgency to address energy sufficiency and environmental quality by reducing carbon emissions and transitioning to carbon-neutral fuels is a major challenge. Accurate and comprehensive data is required to improve our physical understanding for the combustion of novel fuels and support the development of predictive numerical models.

Recent advances in hardware and algorithms have made direct numerical simulations (DNS) a valuable tool in numerical turbulent combustion research \cite{DomingoDNS}, providing a ``model-free" complete description resolved in space and time that cannot be obtained by other means. However, exploiting the full potential of DNS remains a daunting task. The wide range of time and length scales, coupled with the large number of tightly interconnected thermochemical scalars and the complex interactions between flow and chemistry, makes these simulations among the most demanding in computational fluid dynamics (CFD), and necessitate, fast solvers that can efficiently harness the capabilities of the latest supercomputers.

Unfortunately, most combustion research codes are not fully optimized for DNS on the latest supercomputer architectures. Even when they are, they frequently remain limited to simple geometries. Another issue is that most codes are not publicly available, limiting collaboration, transparency and benchmarking. Some of these concerns have been addressed with the release of Pele \cite{Pele}, developed as part of the US Exascale Computing Project. It is an open-source code suite designed for GPU-accelerated systems, employing a second-order finite volume method along with adaptive mesh refinement and an immersed boundary method to handle complex geometries and address real-world applications.

This paper introduces \textit{nekCRF} the open-source successor of LAVp \cite{SKphd, brambilla,behrooz}, specifically designed to align with the recent paradigm shift in HPC towards (GPU) accelerated computing. Drawing on the experience in the development and optimization of its predecessor, \textit{nekCRF} shares the same key algorithmic features. The code supports finite rate chemistry, mixture-averaged transport, conjugate heat transfer, moving meshes and time-varying thermodynamic pressure.

It is built on the CFD code nekRS \cite{nekRS}, which is based on a \verb|MPI+X| programming model supporting CPUs and accelerators through the Open Concurrent Computing Abstraction (OCCA) \cite{occa}. nekRS implements the spectral element method (SEM) offering fast matrix-free operator evaluations with the geometric flexibility of the finite element method \cite{DevilleFischerMund02} enabling the accurate handling of various technical applications in complex geometries. Numerically, the method exhibits far lower levels of numerical diffusion and dissipation at higher polynomial orders at a competitive cost compared to low order method. This makes it a particularly well-suited approximation choice in areas such as DNS of turbulent flows (see e.g. \cite{UQKTH}), where the accurate time-advection of energetic structures such as vortices is a key concern. For optimal performance, the code supports only hexahedral elements, although geometric modeling can be challenging as the generation of accurate, boundary-conforming curvilinear hexahedral meshes remains an open problem. One approach to mitigate this issue is through the use of overset grids which are also supported. Nonconformal adaptive mesh refinement is another potential approach but it still relies on an initial coarse grid, which can be problematic for complex geometries. The highest geometric flexibility is offered by unstructured meshes of simplicial or hybrid elements implemented by solvers like Nektar++ \cite{nektar}. However, it is evident and unavoidable that these element types will face challenges in achieving the same performance as the more naturally tensor-product structured hexahedral elements. Nonetheless, in some cases, the trade-off in performance may be justified. For a detailed description of nekRS development, including the discretization method, algorithmic considerations for the parallel GPU development of nekRS, as well as extensive performance studies on Summit the interested reader can consult \cite{nekRS}. 

The paper is structured as follows.  First, the conservation equations and numerical method are presented. The code is validated in Section \ref{validation} against solutions obtained with well-established codes in setups of increasing complexity. The performance is subsequently studied in detail, including cross-code comparisons in Section \ref{performance}.

\section{Numerical method and code description}\label{numMethod}

The system of equations for reacting flow considered here is based on the low Mach number formulation \cite{ChuKovasznay58,RehmBaum78}. Integrating the low Mach equations demands careful consideration to effectively couple advection with the diffusion and production terms operating on much faster time scales. The coupling strategy becomes a key determinant of the largest timestep size that preserves accuracy, stability, and computational efficiency. Different strategies have been proposed \cite{Nonaka}, each with its own advantages and disadvantages, but discussing them in detail is beyond the scope of this paper. In this work, the algorithm proposed in \cite{TLO97} is employed. 

First, the \textit{thermochemistry} subsystem 

\begin{subequations} \label{eqn:conservationEqsnYT}
\begin{align}
\frac{\partial Y_k}{\partial t} = 
    -\mathbf{v}\cdot\nabla & Y_k +  \frac{1}{\rho} \left(- \nabla\cdot\rho Y_k\mathbf{V}_k + \dot{\omega}_k \right)  \label{eqn:Yk}
\\
  \frac{\partial T}{\partial t}   =
 -\mathbf{v}\cdot\nabla T & + \frac{1}{\rho c_p} \left( \nabla\cdot \lambda\nabla T 
 +\sum_{k=1}^N h_k^0\dot{\omega}_k \right)  \nonumber \\ 
 +\frac{1}{\rho c_p} &\left(- \nabla\cdot \rho T \sum_{k=1}^N c_{p,k}Y_k\mathbf{V}_k + \frac{dp_0}{dt} \right) \label{eqn:T} \\
\left(1-\frac{\mathcal{R}}{c_p} \right) \frac{dp_0}{dt} &= 
\frac{\mathcal{R}}{c_p}\left(\nabla \cdot \lambda \nabla T+\sum_{k=1}^N h_k^0\dot{\omega}_k \right) \nonumber \\ 
-(\nabla \cdot \mathbf{u}) p_0 & - \frac{\mathcal{R}}{c_p} \left(\nabla\cdot \rho T \sum_{k=1}^N c_{p,k}Y_k\mathbf{V}_k  \right) \nonumber \\ 
& + \sum_i \frac{\mathcal{R} T}{W_i} \left(- \nabla\cdot\rho Y_k\mathbf{V}_k + \dot{\omega}_k \right) \label{eqn:dp0th} \\
p_0 &= \rho \mathcal{R} T \label{eqn:idealGas}
\end{align}
\end{subequations}
%
%
is solved, containing the gaseous species mass fractions $Y_k$ and temperature $T$, where $t$ is time, $\mathbf{V}_k, h_k^0, c_{p,k}, \dot{\omega}_k$ the diffusion velocity, enthalpy of formation, heat capacity at constant pressure and net production rate of species $k$, respectively,  $\rho, c_p, \lambda,  p_0$ the mixture density, heat capacity, thermal conductivity, and thermodynamic pressure, and $\mathcal{R}$ the ideal gas constant. The species diffusion velocities $\mathbf{V}_k$ are computed by a mixture-averaged transport model ignoring Soret and Dufour effects
\begin{equation}
    \mathbf{V}_k = -\frac{D_k}{X_k} \nabla X_k + \mathbf{V}_c
\end{equation}
with $X_k$ being the mole fraction and $D_k$ the mixture-average diffusivity of species $k$. A correction velocity $\mathbf{V}_c=-\sum_{k=1}^N Y_k \mathbf{V}_k$ needs to be introduced for mass conservation \cite{CoffeeHeimerl}. 

The system of equations (\ref{eqn:conservationEqsnYT}) is integrated implicitly as a fully-coupled system of nonlinear differential equations using CVODE from the SUNDIALS package \cite{sundials}. The velocity $\mathbf{v}$ in the advection terms is evaluated explicitly using a third-order polynomial extrapolation (EXT3) from previous timesteps. Within the outer (flow) timestep, CVODE selects its own timestep and may advance the thermochemical state using multiple substeps to satisfy the specified relative and absolute tolerance of the solution in time.
To ensure good performance, it is important to set tolerances that are not stricter than the spatial discretization errors. This may require conducting some experiments to determine the appropriate levels. Each internal CVODE step involves Newton iterations with a generalized minimal residual (GMRES) linear solve, where the Jacobian-vector product is approximated by difference quotients. As is typical for finite difference approximations, selecting the optimal perturbation factor can be challenging and significantly affects integrator performance. The approach outlined in \cite{NITSOL} is adopted, which frequently performs better than the default strategy in CVODE. However, even this method is not fully robust and may require additional tuning. To accelerate the linear solve, a customized GMRES solver is used that stores the Krylov basis in reduced (FP32) precision \cite{compressedBasis}. In addition, an alternative classical Gram-Schmidt \cite{MGS} with just one global synchronization per iteration is employed. Krylov-based methods, such as GMRES, often require an effective and fast-to-evaluate preconditioner to perform well, and constructing such a preconditioner is challenging. 
However, the time step size in a DNS is often sufficiently small, and a preconditioner is not necessary to achieve reasonable performance. The right-hand side (RHS) function, evaluated in each Newton step and linear iteration, employs highly optimized kernels for the SEM transport operators, species production rates, and thermodynamic and transport properties. For this purpose, a just-in-time code generator was developed to translate a combustion model expressed in Cantera format \cite{cantera} into platform tuned source code. More details can be found in \cite{nekrk}.

In reactive low Mach number compressible flows, flow and chemistry are tightly coupled through density and viscosity, and the non-zero divergence constraint $Q_T$ is imposed on the velocity
\begin{equation} \label{eqn:continuity} 
\begin{split}
\nabla \cdot \mathbf{v}  &=  Q_T    \\
    Q_T := -\frac{1}{\rho}\frac{D\rho}{Dt} &= \frac{1}{T}\frac{DT}{Dt} 
    +\sum_{k=1}^N\frac{\overline{W}}{W_k}\frac{DY_k}{Dt} 
    -\frac{1}{p_0}\frac{dp_0}{dt}
\end{split}
\end{equation}
\noindent
where $D/Dt$ denotes the material derivative and $\overline{W}$ the mean molecular weight. The momentum equation for the velocity $\mathbf{v}$ is given by
\begin{equation} \label{eqn:conservationEqsnV}
\begin{split}
\frac{\partial \mathbf{v}}{\partial t} &= 
-\mathbf{v}\cdot\nabla\mathbf{v} -\frac{\nabla p_1}{\rho}  \\
&+ \frac{1}{\rho} \left[\nabla\cdot\mu\left( \underline{\underline{\nabla\mathbf{v}}} 
+ (\underline{\underline{\nabla\mathbf{v}}})^T -\frac{2}{3}(\nabla\cdot\mathbf{v})\underline{\underline{\mathbf{I}}}\right) \right]
\end{split}
\end{equation}
where $\mu$ is the dynamic viscosity, $p_1$ the hydrodynamic pressure, $\underline{\underline{\mathbf{I}}}$ the unit tensor, and $^T$ denotes transposition. The \textit{hydrodynamic} subsystem (Eqs.~\ref{eqn:continuity}--\ref{eqn:conservationEqsnV}) is discretized in time using a third-order backward differentiation formula to approximate the time derivative. For the nonlinear convective term, over-integration (dealiasing) is applied in combination with the operator integrating factor splitting method \cite{OIFS}. The resulting semi-discrete coupled system is solved in a three-step procedure \cite{TLO97}: First, an intermediate velocity is evaluated using the explicit contributions. Next, the hydrodynamic pressure is computed to enforce the divergence constraint (Eq.~(\ref{eqn:continuity})). Finally, the velocity is advanced to the next (flow) timestep. Each step is efficiently treated by techniques tailored to the governing physics: classical fourth-order Runge-Kutta for the hyperbolic advection term, diagonally preconditioned conjugate gradient iteration for the viscous block coupled Helmholtz problem, and hybrid multigrid preconditioned GMRES with Chebyshev accelerated Schwarz (polynomial levels) and Jacobi (algebraic levels of coarse p=1 problem) smoothing for the variable coefficient Poisson solve \cite{nekRS}.

\section{Validation} \label{validation}

nekRS has been validated through comparisons with nek5000 and the method of manufactured solutions. Consequently, this section focuses on the reactive low Mach number flow support introduced by \textit{nekCRF}.

\subsection{Homogeneous variable-volume autoignition} 

\begin{figure}
   \centering
   \includegraphics[width=0.5\textwidth]{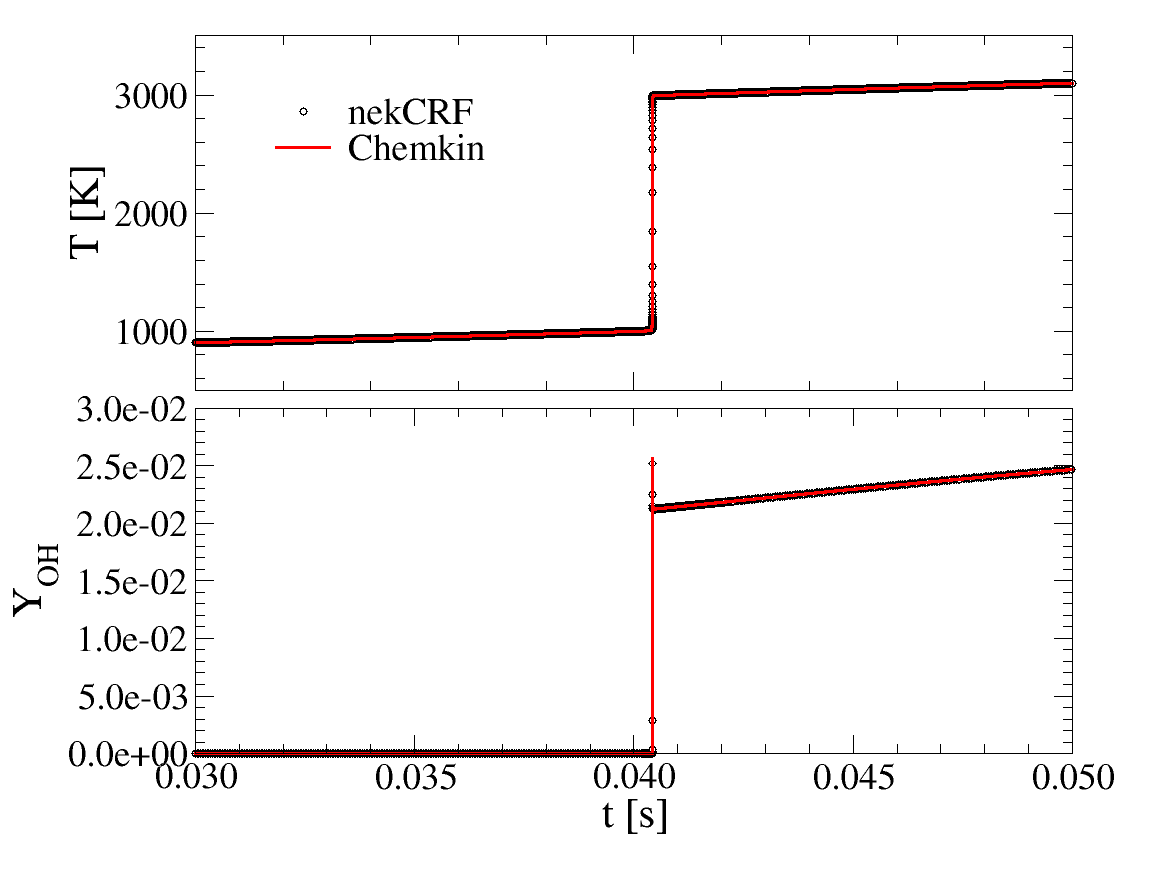}
\caption{Comparison of the time histories of temperature and $Y_{\rm OH}$ (symbols) against the Chemkin solution (lines).}   \label{fig:icen}
\end{figure}

The first test case validates the integration of reaction source terms within a time-varying volume, where the thermodynamic pressure is also subject to variation. A homogeneous stoichiometric H$_2$-air mixture initially at 780~K and 1~atm is compressed in an adiabatic, closed domain mimicking an internal combustion engine of bore 92~mm, stroke 86~mm operating at 500~rpm and compression ratio of 3. The piston kinematics are imposed in the movement of the lower surface in the \textit{nekCRF} model, which tracks the mesh motion using the arbitrary Lagrangian-Eulerian formulation of \cite{Ho1989}. The piston kinematics are also imposed on a variable-volume batch reactor model implemented and solved using Chemkin \cite{CK}. Chemical kinetics are described by the detailed reaction mechanism of Li et al.~\cite{Li}. The temporal evolution of temperature and $Y_{OH}$ of the two solutions are compared in Fig.~\ref{fig:icen}, showing the autoignition resulting from the mixture compression. The relative error in the ignition delay time defined by the time of maximum $Y_{\rm OH}$ with respect to the Chemkin solution is $0.02\%$. 

\subsection{Laminar planar premixed flame}

\begin{figure}
   \centering
   \includegraphics[width=0.5\textwidth]{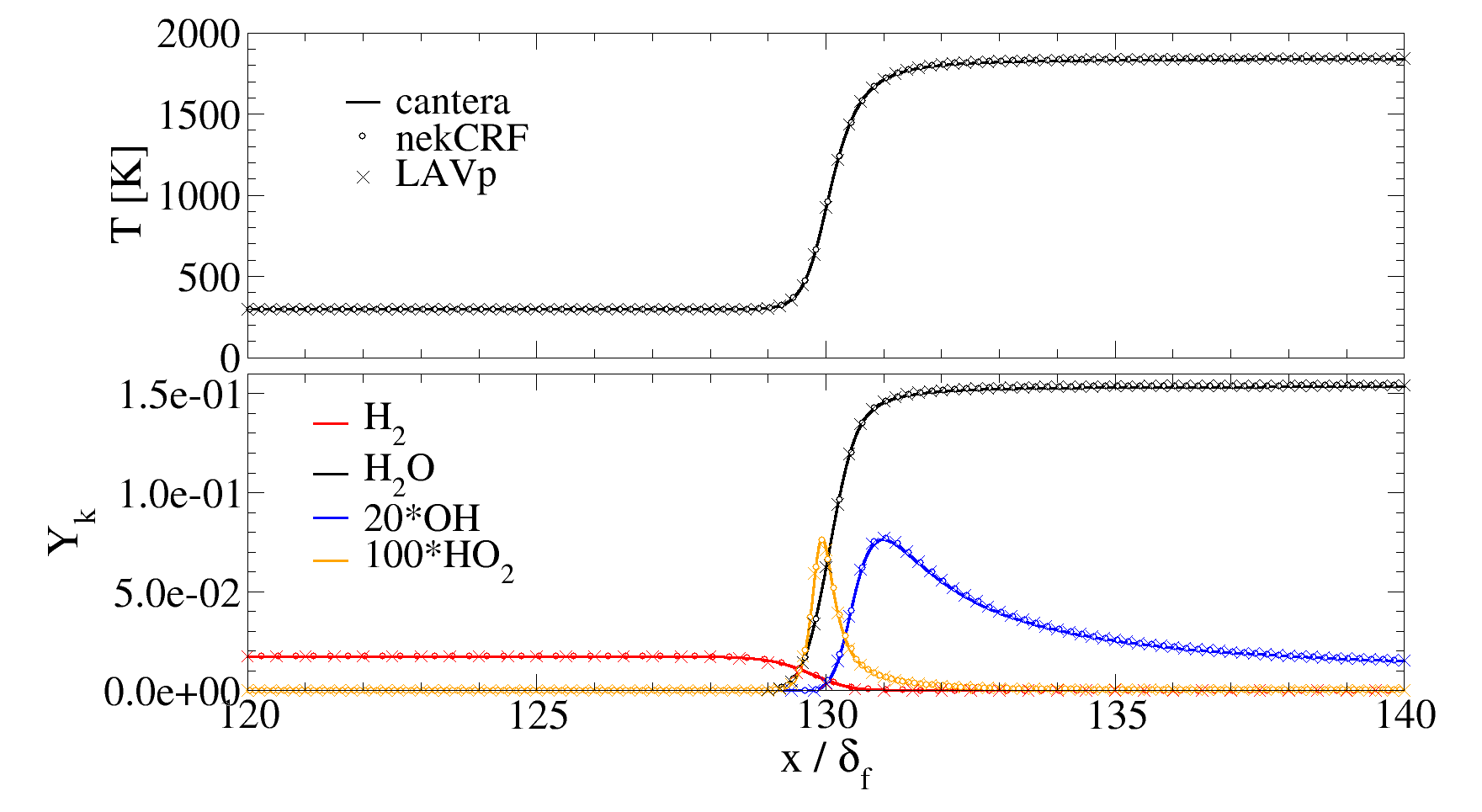}
\caption{Comparison of temperature and selected species mass fraction profiles in the flame normal direction against Cantera (black line) and LAVp ($\times$ symbol) solutions.}   \label{fig:premix}
\end{figure}

In contrast to the previous case, this test includes spatial gradients and incorporates both advection and diffusion terms in addition to the reactive source term. A planar premixed H$_2$-air flame is considered for a lean mixture (equivalence ratio $\phi=0.6$) at $T_u=298$~K and $p=5$~atm. Using a Cantera-based \cite{cantera} freely-propagating, premixed flat flame solver and the Li et al.~\cite{Li} reaction mechanism, the laminar flame speed and thickness are found to be $S_L=51.437$~cm/s and $\delta_f=(T_b-T_u)/\max(dT/dx)=7.516\times10^{-2}$~mm, respectively. The 1-D Cantera solution is interpolated on the spectral element mesh for a domain with length equal to $L=200\delta_f$ and height $H=\delta_f$, where the flame is placed at $x=130\delta_f$. The velocity is set to $U_{in}=S_L$ at the inflow, where Dirichlet boundary conditions (BC) are imposed on temperature and species mass fractions. Zero-Neumann BC are considered at the outflow, while the remaining boundaries are periodic. The discretization uses a non-uniform grid with element of size $h=\delta_f$ around the flame increasing up to $h=25\delta_f$ towards the in- and outflow; the solution on each element is approximated using $p=7^{th}$-order polynomials. Figure~\ref{fig:premix} shows that the flame structure computed with the new solver is in very good agreement with the profiles obtained with Cantera and LAVp.

\subsection{Early flame kernel development}  \label{efkd_0.5M}

\begin{figure}[h]
     \centering
     \begin{subfigure}[b]{0.3\textwidth}
         \includegraphics[width=\textwidth]{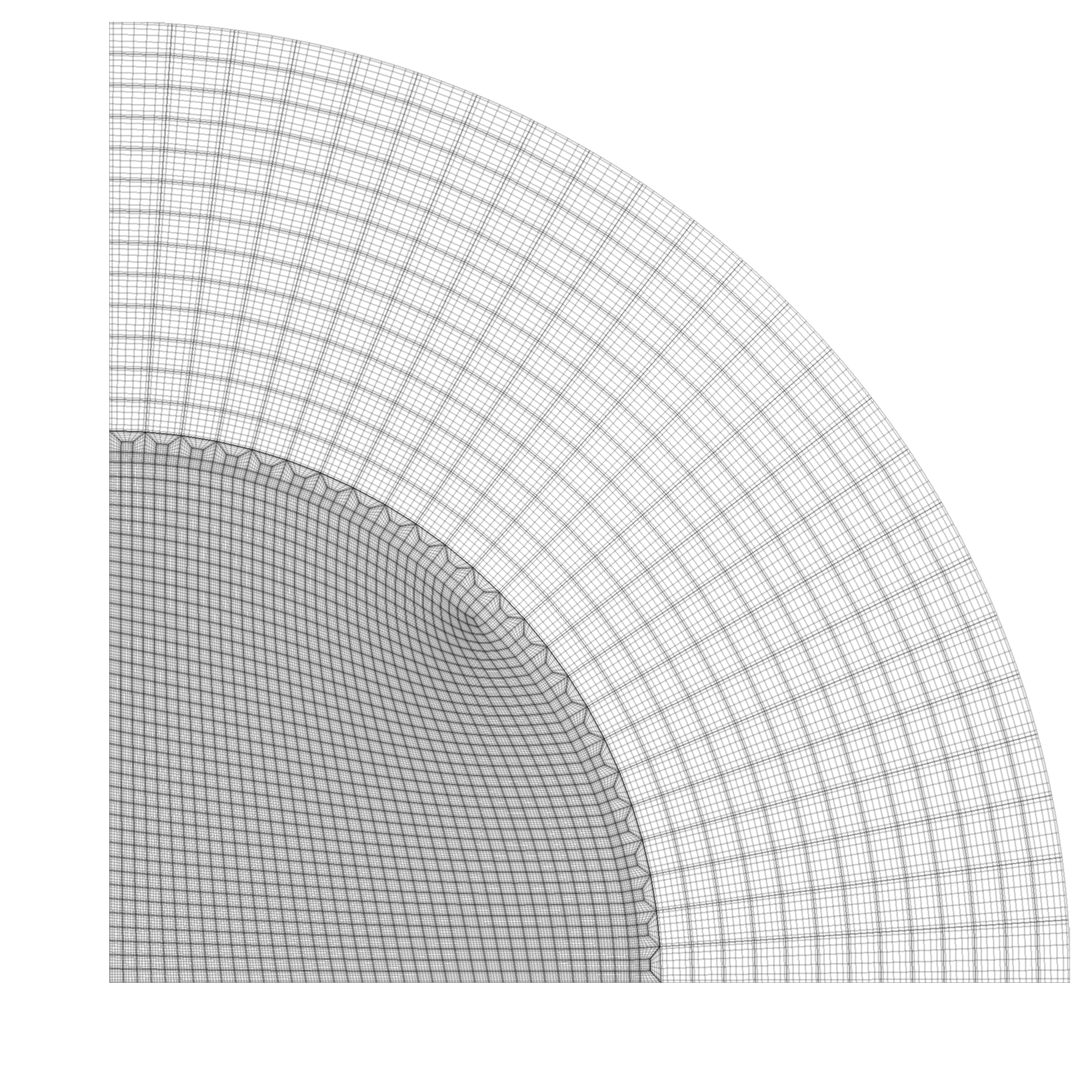}
         \caption{}
     \end{subfigure}
     \hfill
     \begin{subfigure}[b]{0.45\textwidth}
         \includegraphics[width=\textwidth]{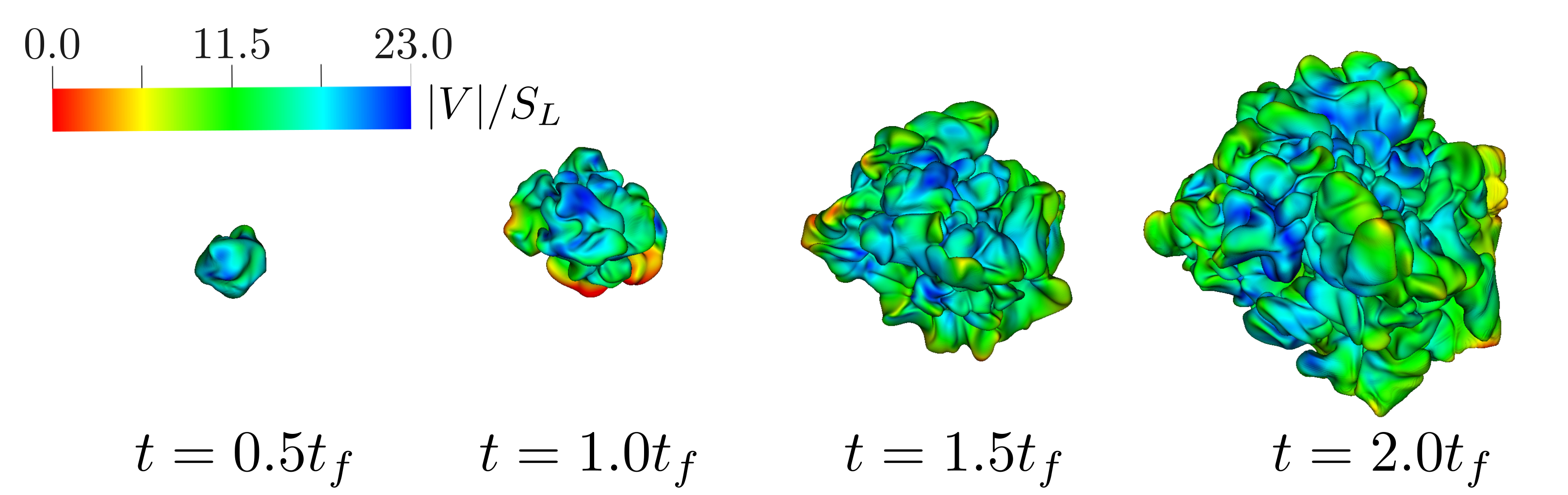}
         \caption{}
     \end{subfigure}
\caption{\label{fig:kernel}(a) Slice of a quarter of the mesh (b) flame kernels defined by the $T=1600$~K isotherm and colored by the flow velocity magnitude at four time instants.}
\end{figure}

Finally, the code is validated in a more complex application by considering a lean ($\phi=0.4$) premixed hydrogen-air mixture initially at temperature $T_0=800$~K and pressure $p=40$~bar. At these conditions, the laminar flame thickness and speed computed with Cantera and the Li et al.~mechanism \cite{Li} are $\delta_f=20$~$\mu$m and $S_L=44.5$~cm/s, respectively, resulting in a flame time $t_f=\delta_f/S_L=4.55\times10^{-5}$~s. The initial turbulent flow field is homogeneous and isotropic with integral length scale $l_I = 15.1\delta_f$ and turbulent intensity $u'=6.6S_L$, and was constructed following the methodology proposed in \cite{Falkenstein}, where the homogeneous and isotropic turbulent fields were generated with the controlled linear forcing method described in \cite{Bassenne}. Ignition of the mixture is achieved with a centrally-located energy deposition source varying smoothly in space and time as discussed in \cite{Falkenstein}. A spherical mesh with a diameter of diameter $D=144\delta_f$ is used (Fig.~\ref{fig:kernel}(a)) comprising $E=0.5$~M spectral elements  with a polynomial order $p=7$, resulting in 2.23 billion degrees of freedom (DOFs), taking into account the 13 unknowns per grid point. The excess species $N_2$ is calculated as $Y_{N_2} = 1 - \sum_{i \neq N_2} Y_i$. In the central region of radius $R\le41\delta_f$, the elements are of size $\delta_f$ in all directions, while in the outer region they are of size $2.4\delta_f$ in the radial direction and increase from $2.8\delta_f$ to $4.8\delta_f$ in the other directions. 

\begin{figure}[h]
   \centering
   \includegraphics[width=0.48\textwidth]{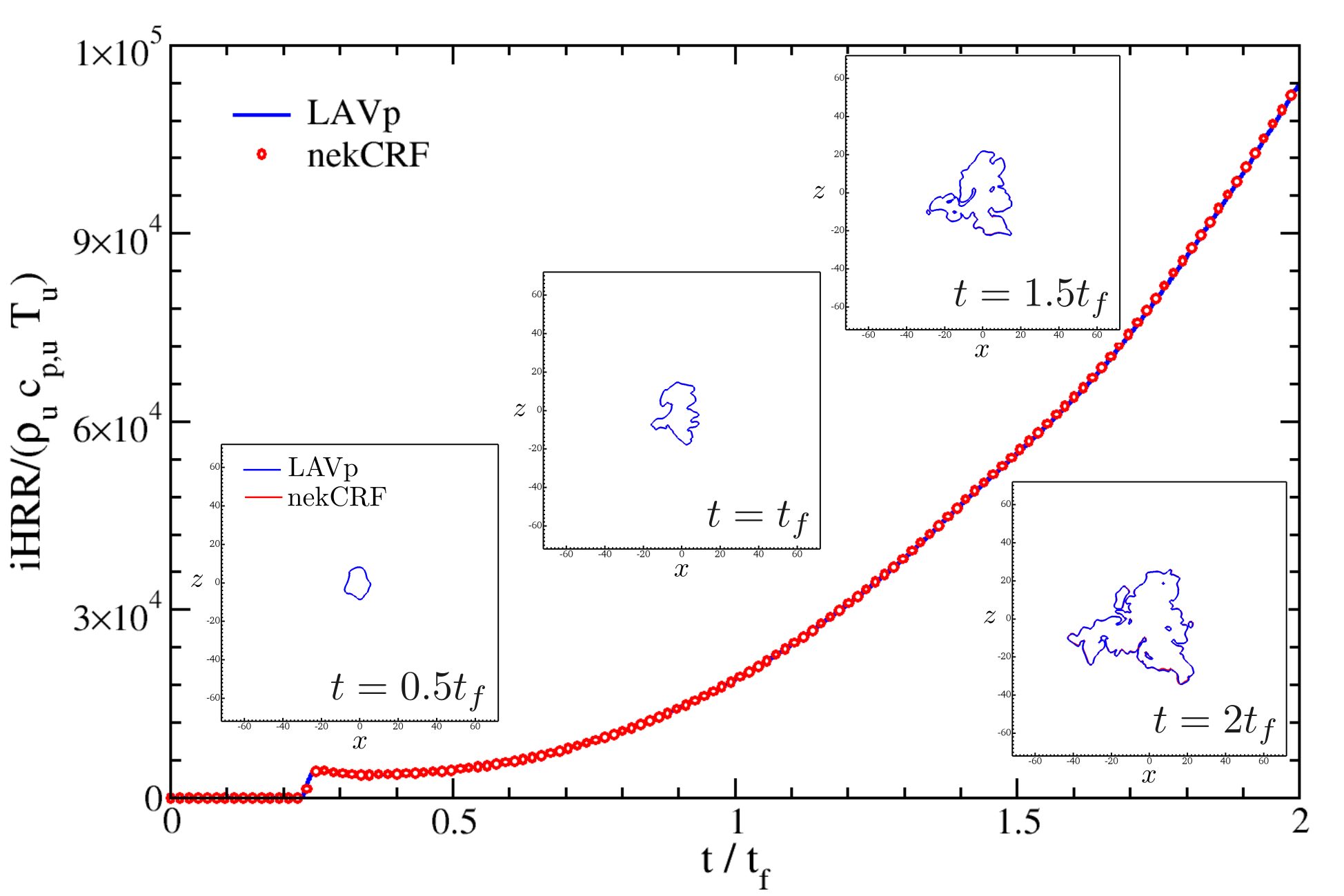}
\caption{Comparison of the time histories of the integral heat release rate and instantaneous isocontours of temperature. Insets:  $T=1600$K isotherms on an $x-z$ slice at four time instants.}   \label{fig:EFKD_comp}
\end{figure}

CVODE absolute tolerances were set to $10^{-4}$ of $\max\{Y_k\}$ of the planar flame, while the relative tolerance was $10^{-6}$ for all variables. For the linear solvers in the fluid solve, the absolute residual tolerance for pressure and velocity were $10^{-5}$ and $10^{-7}$, respectively.
The addition of heat during the initial $0.3t_f$ leads to the establishment of a flame kernel from which a propagating flame front is generated that interacts with the decaying turbulent flow field (Fig.~\ref{fig:kernel}(b)). Figure~\ref{fig:EFKD_comp} shows good agreement in the evolution of the volumetric integral of the heat release rate, $iHRR=\int_V \sum_k h_k \dot{\omega}_k dV$, and the instantaneous isocontours of temperature with respect to LAVp for times up to $2t_f$ when the flame remains within the high resolution region.
 
\section{Performance} \label{performance}

In this section, the performance is initially assessed by comparing \textit{nekCRF} with LAVp. Subsequently the strong scaling capabilities and compute kernels of \textit{nekCRF} are explored. Unless specified otherwise, the performance experiments were conducted on JUWELS Booster at the J\"{u}lich Supercomputing Centre (JSC). The system consists of 936 compute nodes, each equipped with 4 NVIDIA A100 SXM4 GPUs.

It is noteworthy that the reported thermochemistry timings are approximately twice as fast compared to running CVODE with default settings and without the customized Jacobian-vector product and linear solver modifications discussed in Section~\ref{numMethod}. This speedup is primarily due to using a difference quotient perturbation factor about 100 times smaller than the default and a relaxed linear convergence safety factor of 0.5 instead of 0.05 for each Newton iteration.

\subsection{Comparison to LAVp}

To compare the performance, the case presented in Section~\ref{efkd_0.5M} is evaluated using LAVp on the LUMI-C HPC system at the CSC IT Center for Science in Finland. It is important to recognize that the reported speedup factors are generally influenced by specific implementation details, hardware setup, and problem size. Initial microbenchmarks provide valuable insights into expected behavior, showing that reaction rates on GPUs can achieve speedups between 6x and 9x, depending on the size of the chemistry model. Other performance-relevant kernels, which are primarily limited by memory bandwidth, can achieve peak speedups of 9x. This performance difference aligns with the hardware capabilities. However, in typical production runs, the overall speedup tends to be lower because certain parts of the workload fit within the CPU's last-level cache, which provides more competitive bandwidth compared to the GPU's memory bandwidth.

Striving for a closely matching comparison, it is important to note that \textit{nekCRF} has been more thoroughly optimized compared to LAVp, which may lead to an overestimation of performance improvements. To assess potential speedup ranges for the entire solver, multiple experiments were conducted and the results are summarized in Table \ref{tab:SpeedUpComparison}. 

\begin{table}[h]
\centering
\begin{tabular}{|c|c|}
\hline
        use case & speedup  \\  \hline
        $S_{max}$  & 22 \\
        $R_{max}$  & 14 \\
        $R_{0.5}$  & 4\\ 
\hline
\end{tabular}
\caption{Speedup factors compared to LAVp.} \label{tab:SpeedUpComparison}
\end{table}

The maximum achievable speedup per compute unit (CU) was measured to be $S_{max}=22$. In this scenario, both codes handle a large local problem size, which minimizes the impact of caching effects on the CPU. However, when considering the maximum throughput $R_{\text{max}} = 1/(nCU \cdot t)$ of each solver, where $nCU$ is the number of compute units (24 GPUs for \textit{nekCRF} and 300 CPUs using all available cores for LAVp), the speedup reduces to 14 due to caching effects. Notably, $R_{0.5}$, the speedup at 50\% of $R_{\text{max}}$, utilizing 480 GPUs on JUWELS Booster and 2000 CPUs on LUMI, offers a different perspective. This strong scaling use case demonstrates that while \textit{nekCRF} remains faster on a GPU-based system, its competitive advantage diminishes as communication overhead becomes more significant.

\subsection{Strong scaling}
To stress test strong scaling, the investigation focuses on the configuration outlined in Section~\ref{efkd_0.5M}, employing a refined mesh with approximately doubled resolution in each spatial direction and a two times smaller outer timestep. The refinement leads to a case with $E=4.1$ million spectral elements and 18 billion DOFs. Computations are conducted across a range of N=50 to 800 compute nodes (85\% of the JSC system size), with the lower limit determined by the available GPU memory. However, the focus here is on up to 600 nodes, as the parallel efficiency $\eta$ for larger node counts becomes smaller than $50\%$. 

This scenario poses a challenge for strong scaling due to the relatively small size of the hydrogen reaction mechanism, which involves only 9 species participating in 21 reactions. Larger chemistry models typically require more local computations, leading to improved strong scaling of the thermochemistry solve. They also reduce the relative cost of the fluid solve, which is much harder to strong scale due to the smaller system size.

The solver statistics averages over the outer (fluid) timestep are reported in Table~\ref{tab:EFKD_solverStats} including pressure iterations (pIter), velocity iterations (vIter), CVODE timesteps (cvSteps), Jacobian-vector evaluation per CVODE step (jtv/cvSteps), non-linear Newton iterations per CVODE step (nni/cvSteps), and ratio of linear to nonlinear iterations (nli/nni). The counters cvSteps and Jtv/cvSteps provide a rough measure of the overall cost of the thermochemistry solve. The ratio nni/cvSteps measures the performance of the nonlinear solver (typical values ranging from 1.1 to 1.8 according to \cite{cvodeDocumentation}), while the ratio nli/nni measures the performance of the Krylov linear solver, and thus (indirectly) the condition number of the linear system. 

\begin{table}[h]
\centering
 \begin{tabular}{|l|c|}
\hline
pIter & 11  \\
vIter & 34 \\
cvSteps & 4.5 \\
jtv/cvSteps & 7.4  \\
nni/cvSteps & 1.1 \\
nli/nni &  6.4 \\ \hline
\end{tabular}
\caption{Solver statistics averaged over 200 timesteps.} \label{tab:EFKD_solverStats}
\end{table}

\begin{figure}
\centering
 \includegraphics[width=0.5\textwidth]{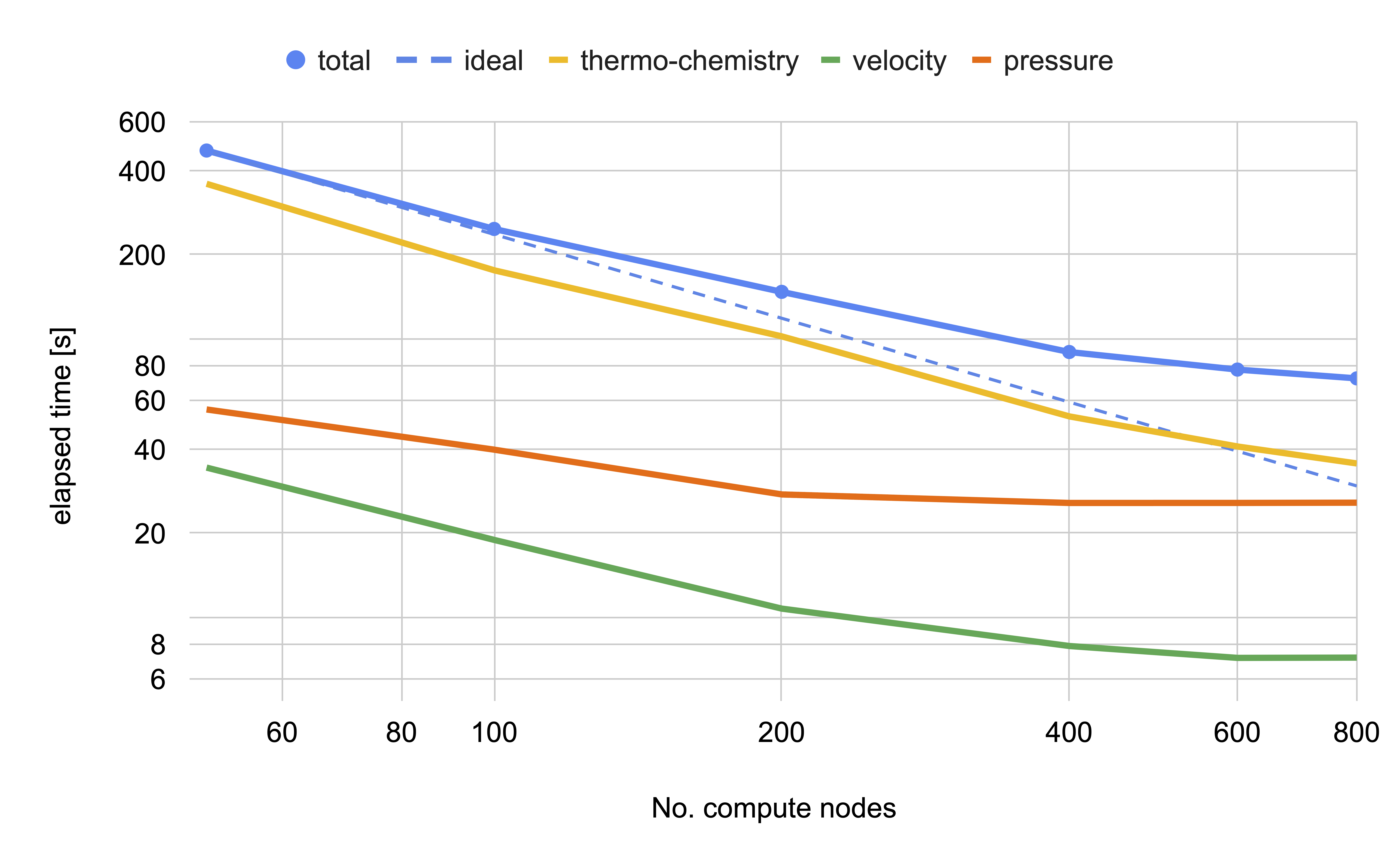}
\caption{\label{fig:EFKD_scaling}Strong scalability of the different solver components.}
\end{figure}

The  elapsed time for the integration of 200 timesteps with $dt=10^{-3}t_f$ ($CFL \approx 3$) is shown in Fig.~\ref{fig:EFKD_scaling} together with its breakdown into the contributions of the velocity, pressure and thermochemistry parts. The solvers for velocity, pressure, and thermochemistry exhibit different scaling characteristics. Among them, the thermochemistry solver is the most compute-intensive, with $\eta_{tc}=0.72$ on 600 compute nodes. On the other hand, the fluid solver shows diminishing returns (poor scaling) beyond 200 nodes, specifically when the number of grid points per processing unit $n=1.75M$ is considerably smaller than $n_{0.8}$ - the number of grid points per GPU required to achieve $\eta=0.8$. According to \cite{nekRSScaling}, the value of $n_{0.8}$ was measured to be around 4-5M on ANL's Polaris, a system similar to JUWELS Booster. This results in an overall parallel efficiency of $\eta=0.51$ on 600 nodes. It is noteworthy that in typical production runs employing a more complex chemistry model and/or higher number of grid points per GPU, the efficiency will be significantly higher.

\begin{table}
    \setlength{\tabcolsep}{5pt} 
    \centering
    \begin{tabular}{l|rrrr}
                    & N=50& $\%$ & N=600 & $\%$ \\ \hline
     total          & 339s &  & 39.4s &  \\ 
     newton solve   & 64.4s & 19 & 8.3s & 21 \\
     linear solve   & 274.6s & 81 & 31.1s & 79 \\
     \\
     \textit{low-level components} \\
     other & 98.2s  & 29 & 10.0s & 25  \\
     diffusion      & 95.8s & 28 & 8.15s & 21 \\
     gramSchmidt    & 57.2s & 17 & 5.60s & 14 \\
     MPI\_Allreduce & 2.60s & 1 & 4.18s & 11  \\
     MPI halo& 12.8s & 4 & 3.40s & 9 \\
     local$QQ^T$    & 20.1s & 6 & 3.30s & 8 \\
     MPI pack/unpack &  8.32s & 2 & 2.31s & 6 \\
     advection      & 25.4s & 7 & 2.54s & 6 \\
     rates          & 22.9s & 7 & 2.23s & 6\\
     properties     & 6.32s & 2 & 0.75s & 2 \\
    \end{tabular}
    \caption{Time breakdown of the thermochemistry solve.}
    \label{tab:EFKD_breakdown}
\end{table}

To delve into the details of the thermochemistry solver, a time breakdown is presented in Table~\ref{tab:EFKD_breakdown}, revealing a significant influence of the embedded linear solve at each Newton step, consuming roughly 80\% of the overall execution time. Besides the Jacobian-vector product evaluation (using a difference quotient of the RHS) in each GMRES iteration, the local Gram-Schmidt process accounts for up to 17\% of total cost on average. The diffusion operator emerges as the most time-consuming kernel within the RHS. As expected for this small chemistry model, the computation of reaction rates and properties accounts for only a small fraction (8\%) of the total time. The ``other" category primarily consists of various Level 1 BLAS type operations.

To further analyze the observed parallel efficiency of the thermochemistry solve, it is useful to separate it into two factors: computational efficiency and communication efficiency. 

Computational efficiency measures how well local compute kernels scale relative to peak performance and was evaluated at around 0.9 across all kernels. The efficiency of individual kernels varies depending on the kernel type. Kernels of the gather/scatter type (local$QQ^T$ and MPI pack/unpack) exhibit less favorable scaling characteristics compared to all other kernels. These kernels tend to be short (for more details see Section~\ref{efficiency}), making latency a limiting factor. However, although the relative cost rises, it remains within a moderate range below $14\%$.

Communication efficiency (computation time to total runtime ratio) measures the overhead caused by off-device communication as the problem is scaled. The main limiter in the present case is associated with inner products (in the linear solver) necessitating global synchronization, which are known to scale poorly due to internode latency. To mitigate the impact, a Gram-Schmidt variant is employed that requires only one global synchronization step \cite{MGS}. Yet, inner products to compute the initial and final residual norms in the Newton or linear solver are challenging to reduce or overlap, posing limitations on scalability. This is evident in the increasing relative \verb|MPI_Allreduce| cost. The global $C^0$ assembly is imposed through a nearest-neighbor exchange locally ($QQ^T$) and across neighbor processes (MPI halo). On average, messages of size 212 kB are exchanged among 17 neighbors  achieving an aggregated bidirectional bandwidth of 148 GB/s on 50 nodes (74\% of theoretical node injection bandwidth 200 GB/s), which decreases by 17\% on 600 nodes. However, the off-device communication (\verb|MPI_Neighbor_alltoall|) within $QQ^T$ can be fully overlapped by local computation, allowing the effective cost to be hidden to a large extent. Consequently, the most significant factor contributing to the effective communication cost is the \verb|MPI_Allreduce| operation. The measured communication efficiency was approximately 0.8. 


\subsection{Compute kernels}\label{efficiency}

This section aims to assess the kernel performance, with a focus on the top five most time-consuming kernels of the thermochemistry solver, as summarized in Table \ref{tab:knlBW}. The evaluation includes identifying the dominant performance limiter and presenting a figure of merit (FOM).

\begin{table}
\centering
 \begin{tabular}{|l|l||l|}
\hline
kernel & limiter & FOM  \\  \hline
diffusion & GMEM bw & 987 GB/s \\
gramSchmidt & GMEM bw & 802 GB/s \\
rates & latency & 2.4 TFLOPS  \\
local$QQ^T$ & GMEM bw & 580 GB/s \\
advection & GMEM bw & 935 GB/s  \\
\hline
\end{tabular}
\caption{Performance top 5 kernels.} \label{tab:knlBW}
\end{table}

The data was collected on 50 compute nodes (Tab.~\ref{tab:EFKD_breakdown}) with the local workload sufficiently high to achieve peak kernel throughputs. Kernels in the other category achieve an effective memory bandwidth close to the sustainable memory bandwidth of 1370GB/s \cite{STREAM} as expected for a simple streaming type. Most of the top 5 kernels are primarily limited by global memory bandwidth (GMEM bw), with other factors also constraining performance and preventing the attainment of ideal streaming-type performance. The measured effective memory bandwidth for these five most time-consuming kernels ranges from 42\% to 73\%. The gather/scatter type operations pose a challenge, as they exclusively operate on the element or partition surface, resulting in short kernels with irregular memory access patterns that waste memory bandwidth as the element data layout is lexicographical ordered to minimize indirect addressing in the local tensor contractions. SEM operators, such as advection and diffusion, as well as the Gram-Schmidt kernels, utilize shared resources (registers and scratch pad memory) to cache input data or intermediate results. This greatly reduces the amount of GMEM data transferred and improves the overall performance. However, it may lead to lower occupancy and, consequently, a decrease in achievable bandwidth. Optimizing the reaction rate kernel is challenging due to large working sets, irregular execution orders, and complex data access patterns, with latency often becoming a significant constraint. Several GPU-optimized implementations have been developed, including those in \cite{singe, TChem, ratesML, Pele}. In the absence of a clear performance roofline, the floating-point operations per second (FLOPS) is used as a performance metric. Further details and performance comparisons are discussed in \cite{nekrk}.

\section{Conclusions and outlook}

\textit{nekCRF} was developed to enable efficient direct numerical simulations of reactive low Mach flows on accelerator-based supercomputers. The solver supports unstructured curvilinear boundary conforming meshes for accurate representation of complex geometries, finite rate chemistry, mixture-averaged transport, conjugate heat transfer, moving mesh capabilities and time-varying thermodynamic pressure variation. 
Validation was based on comparison with established numerical solvers and various reactive use cases. During development, the initial focus was on NVIDIA GPUs, for which computationally efficient chemistry kernels were implemented. The strong scalability of \textit{nekCRF} was tested on up to 3600 GPUs on JUWELS Booster, showing that it scales efficiently. Additionally, cross-code performance comparisons demonstrated that \textit{nekCRF} outperforms its predecessor, LAVp.

Ongoing efforts involve incorporating support for AMD and Intel GPUs, prioritizing performance optimization for these architectures, and ultimately testing our solver on real exascale use cases. This may prompt a reevaluation of the thermochemistry solver to enhance performance when dealing with more severe stiffness or incorporating additional multiphysics aspects like radiation, soot, or heterogeneous reactions. A viable approach could involve leveraging methodologies akin to those outlined in \cite{sdc}. Moreover, mixed-precision techniques are being investigated to further accelerate the solver. Finally, evaluating against other well-established solvers will help to clarify the competitiveness of the performance.

\section{Availability}

At the time of writing, the code has not been made publicly available. However, after a testing phase, it will be released as open-source under the BSD license and can be accessed on GitHub: \newline
\url{https://github.com/nekCRF}.

\section{CRediT authorship contribution statement}

\textbf{Stefan Kerkemeier}: Software, Methodology, Investigation, Writing. 
\textbf{Christos E. Frouzakis}: Writing, Validation, Funding acquisition. 
\textbf{Ananias G. Tomboulides}: Review, Supervision. 
\textbf{Paul Fischer}: Review. 
\textbf{Mathis Bode}: Review, Funding acquisition, Supervision.

\section{Acknowledgements}
This project received funding from the European Union’s Horizon 2020 research and innovation program under the Center of Excellence in Combustion (CoEC) project, grant agreement No 952181. The authors gratefully acknowledge the Gauss Centre for Supercomputing e.V. 
for funding this project by providing computing time on the GCS Supercomputer JUWELS at J\"{u}lich Supercomputing Centre (JSC). Special thanks to Peng Wang from NVIDIA for valuable assistance in fine tuning some critical kernels.

\bibliographystyle{unsrt}
\bibliography{refs}

\end{document}